\documentclass{PoS}

\usepackage{epsfig}

\usepackage{multicol}
\usepackage{rotating}

\include{symbolsJHEP}

\title{New physics reach of CP violating observables in the decay  $\bar B_d  \rightarrow \bar K^* \ell^+ \ell^-$}

\ShortTitle{ New physics reach of CP violating observables in $\bar B_d \rightarrow \bar K^* \ell^+ \ell^-$}

\author{Ulrik Egede \\
	Imperial College London, London SW7~2AZ, United Kingdom\\
	E-mail: \email{U.Egede@imperial.ac.uk}}
\author{\speaker{Tobias Hurth}\\
       \\ CERN, Dept. of Physics, Theory  Division, CH-1211 Geneva 23, Switzerland\\
       Institute for Physics, Johannes Gutenberg-University, D-55099 Mainz, Germany\\
           E-mail: \email{tobias.hurth@cern.ch}}
\author{Joaquim Matias \\
	IFAE, Universitat Aut\`onoma de Barcelona, 08193 Bellaterra, 
Barcelona, Spain\\
	E-mail: \email{matias@ifae.es}}
\author{Marc Ramon\\
	IFAE, Universitat Aut\`onoma de Barcelona, 08193 Bellaterra, 
Barcelona, Spain\\
	E-mail: \email{mramon@ifae.es}}
\author{Will Reece \\
	Imperial College London, London SW7~2AZ, United Kingdom\\
	E-mail: \email{w.reece06@imperial.ac.uk}}

\abstract{ 
We discuss theoretical and experimental preparations for an indirect new physics  search 
using the rare decay  $\bar B_d \rightarrow \bar K^{*0} (\rightarrow K \pi)  \mu^+\mu^-$ focusing on 
CP violating observables.  
The separation of new physics effects and hadronic uncertainties is the key issue 
 when using flavour observables in a  new-physics search.
Our analysis is based on QCD factorization and soft-collinear effective theory  and critically examines the new physics
reach of those observables via  a detailed 
error analysis due to scale dependences, 
form factors, and other input parameters;  we also explore the experimental sensitivities at LHCb 
using a full-angular fit method; finally, we make the impact of the unknown $\Lambda/m_b$ 
corrections manifest in our theoretical predictions. 
}

\FullConference{European Physical Society Europhysics Conference on High Energy Physics,
EPS-HEP 2009,\\
		 July 16 - 22 2009\\
		 Krakow, Poland\\
	\vspace{0.5cm}
	CERN-PH-TH/2009-208, MZ-TH/09-45, IC/HEP/09-15, UAB-FT-674}

\begin{document}

\section{Introduction}

At the beginning of the LHC era and
close to the end of the $B$ factories   at SLAC \cite{BABAR}
and at KEK \cite{BELLE} and of the Tevatron $B$ physics experiments
\cite{TevatronB1,TevatronB2}, 
all experimental data
on flavour mixing and  CP violating phenomena are consistent with the simple CKM theory
of the standard model (SM) \cite{CERNFLAVOUR2}, which means   that 
all flavour-violating processes between quarks are governed by a $3 \times 3$ 
unitarity matrix, usually referred to as  Cabibbo-Kobayashi-Maskawa (CKM)
matrix \cite{CKM}. 
The CKM matrix is fully described by four real parameters, three
rotation angles and one phase. It is this phase that
represents the only source of CP violation in the SM and that 
now allows for an unified description of all the CP violating
phenomena.  This is an impressive  success of the SM and the 
CKM theory and 
can be illustrated by the overconstrained  triangles in
 the complex plane which reflect the unitarity of the 
CKM matrix. The successful theory  was honored by last year's nobel prize
in physics \cite{NOBEL}. 
Thus,  the CKM mechanism is the dominating effect for CP
violation and flavour mixing in the quark sector; however, there is still room for sizable new effects and new flavour structures because 
the flavour sector has only been tested  at the $10\%$ level especially in the $b \rightarrow s$ sector. 
In particular, CP violating observables are a good testing ground for new physics scenarios. While 
the SM is very predictive by describing all CP violating phenomena via one parameter, many new 
physics models offer many new CP phases.

In Ref.~\cite{Egede:2008uy,Egedeetal1b},  we worked out   the theoretical and experimental preparations for an
indirect new physics  (NP) search using the rare decay   $\bar B_d  \rightarrow \bar K^* \ell^+ \ell^-$                    based on the QCDf/SCET approach:  QCD corrections are included at the next-to-leading order level
and also the impact of the unknown $\Lambda/m_b$ corrections is  made explicit. 
The full angular analysis of the decay $\bar B_d  \rightarrow \bar K^* (\rightarrow K^- \pi^+) \ell^+ \ell^-$
at the LHCb experiment 
offers great opportunities for the 
new physics search.  New observables can be designed to be sensitive to a specific 
kind of NP operator within the model-independent analysis using the effective field theory approach.
The new observables  $A_T^{(2)}$, $A_T^{(3)}$, and $A_T^{(4)}$  are shown to be highly sensitive to right-handed  currents.  Moreover, it was  shown that the previously discussed angular distribution
$A_T^{(1)}$  cannot be measured at either LHCb  or at a Super-$B$ factory.

In the present letter we extend this preparation work to CP violating observables in the 
rare decay. We already anticipate here that,  in contrast to claims in the literature, the new physics
reach of such CP violating observables is rather limited. 
More  details of our analysis with further results will be published in a forthcoming 
paper~\cite{Egedeetal2}. 

\section{CP asymmetries}

The decay\,  \BdbKsll with $\Kstarzb \to \Km \pip$ on the mass shell  is
completely described by four independent kinematic variables, the lepton-pair
invariant mass squared, \qsq, and the three angles $\theta_l$, $\theta_{K}$,
$\phi$. Summing over the spins of the final particles, the differential decay
distribution of\,  \BdbKsll can be written as
\begin{equation}
\label{differential decay rate}
  \frac{d^4\Gamma}{dq^2\, d\cos\theta_l\, d\cos\theta_{K^*}\, d\phi} =
   \frac{9}{32\pi} J(q^2, \theta_l, \theta_{K^*}, \phi)\,,
\end{equation}
The dependence on the three angles can be made more explicit:
\vspace{2cm}
\begin{eqnarray} 
  J(q^2, \theta_l, \theta_K, \phi)=&  
      J_1^s \sin^2\theta_K + J_1^c \cos^2\theta_K
      + (J_2^s \sin^2\theta_K + J_2^c \cos^2\theta_K) \cos 2\theta_l + J_3 \sin^2\theta_K \sin^2\theta_l \cos 2\phi 
\nonumber \\       
    & + J_4 \sin 2\theta_K \sin 2\theta_l \cos\phi  + J_5 \sin 2\theta_K \sin\theta_l \cos\phi+ (J_6^s \sin^2\theta_K +  {J_6^c \cos^2\theta_K})  \cos\theta_l 
\nonumber \\      
    & + J_7 \sin 2\theta_K \sin\theta_l \sin\phi  + J_8 \sin 2\theta_K \sin 2\theta_l \sin\phi + J_9 \sin^2\theta_K \sin^2\theta_l \sin 2\phi\,.
\end{eqnarray}
The angles are defined in the intervals
\begin{equation}
  \label{int:region}
  -1\leqslant\cos\theta_l\leqslant 1\, ,\qquad
  -1\leqslant\cos\theta_{K}\leqslant 1\, , \qquad
  -\pi\leqslant\phi < \pi\, ,
\end{equation}
where in particular it should be noted that the $\phi$ angle is signed.
The corresponding decay rate for the CP conjugated decay mode $B^0 \rightarrow K^{*0} (\rightarrow K^+ \pi^-) \mu^+\mu^-$ is given by 
\begin{equation}
  \frac{d^4 \bar{\Gamma}}{dq^2\, d\cos\theta_l\, d\cos\theta_{K^*}\, d\phi} =
   \frac{9}{32\pi} \bar{J}(q^2, \theta_l, \theta_{K^*}, \phi)\,.
\end{equation}
As shown in \cite{Krueger1}, the corresponding functions 
$\bar{J}_i( q^2, \theta_l, \theta_K, \phi)$  are connected to functions $J_i$ in the following way:
\begin{equation} 
J_{1,2,3,4,7} \rightarrow \bar{J}_{1,2,3,4,7}, \,\,\,\,\, J_{5,6,8,9} \rightarrow - \bar{J}_{5,6,8,9}
\label{Jtrafos}
\end{equation}
where $\bar{J}_i$ equals $J_i$ with all weak phases conjugated.

The $J_i$ depend on products of the seven complex $K^*$ spin amplitudes,
$A_{\bot L/R}$, $A_{\| L/R}$, $A_{0 L/R}$, $A_t$  with each of these a function of 
$q^2$. 
The amplitudes are just linear combinations of the well-known helicity amplitudes 
describing  the $B\to K\pi$ transition. They  
 can be parameterised in terms of the seven $B\to K^*$  
form factors by means of a narrow-width approximation. They also depend 
on the  short-distance Wilson coefficients $C_i$ corresponding to the various
operators of the effective electroweak Hamiltonian.  The precise definitions of 
the form factors and of the effective operators  are given in Refs.~\cite{Egede:2008uy,Egedeetal2}. 
Assuming only the three most important SM operators for this decay  mode, namely 
${O}_7$,   ${O}_9$, and      ${O}_{10}$, and the chirally flipped ones, 
 being numerically  relevant   we have
\begin{eqnarray} 
A_{\perp L,R} &=& N \sqrt{2} \lambda^{1/2} \bigg[ 
\{ (C_9^{eff} + C_9^{eff \prime}) \mp (C_{10}^{eff} + C_{10}^{eff  \prime})\} \frac{ V(q^2) }{ m_B + m_\kstar} 
 + \frac{2m_b}{q^2} (C_7^{eff} + C_7^{eff \prime}) T_1(q^2)
\bigg] \nonumber\\
A_{\parallel L,R} &=& - N \sqrt{2}(m_B^2 - m_\kstar^2) \bigg[ \{ (C_9^{eff} - C_9^{eff \prime}) \mp (C_{10}^{eff} - C_{10}^{eff \prime}) \}
\frac{A_1(q^2)}{m_B-m_\kstar}       \nonumber\\
& & +\frac{2 m_b}{q^2} (C_7^{eff} - C_7^{eff\prime}) T_2(q^2)
\bigg] \nonumber \\
A_{0L,R} &=& - N/ (2 m_\kstar \sqrt{q^2})  \,\,  \bigg[ \{ (C_9^{eff} - C_9^{eff\prime}) \mp (C_{10}^{eff} - C_{10}^{eff\prime}) \} \times  \nonumber\\   
& & \times \{ (m_B^2 - m_\kstar^2 - q^2) ( m_B + m_\kstar) A_1(q^2) 
 -\lambda A_2(q^2) / (m_B + m_\kstar)
\} 
\nonumber\\
& &  + {2 m_b}(C_7^{eff} - C_7^{eff\prime}) \{
 (m_B^2 + 3 m_\kstar^2 - q^2) T_2(q^2)
-\frac{\lambda}{m_B^2 - m_\kstar^2} T_3(q^2) \}
\bigg]\nonumber \\
 A_t &=&  N   \lambda^{1/2}   / \sqrt{q^2}   \{ 2 (C_{10}^{eff} - C_{10}^{eff\prime}) \} A_0(q^2) 
\label{SCETKspin}
\end{eqnarray} 
where the $C_i$ denote the corresponding Wilson coeficients and 
\begin{equation}
  \label{eq:Lambdadef}
  \lambda= m_B^4  + m_{K^*}^4 + q^4 - 2 (m_B^2 m_{K^*}^2+ m_{K^*}^2 \qsq  + m_B^2 \qsq),
\end{equation}
\begin{equation}
N=\sqrt{\frac{G_F^2 \a^2}{3\cdot 2^{10}\pi^5 m_B^3}
|V_{tb}^{}V_{ts}^{\ast}|^2 \qsq \lambda^{1/2}
\sqrt{1-\frac{4 m_l^2}{\qsq}}}.
\end{equation}
In Refs.~\cite{Krueger1,Krueger2},  it was shown that eight CP-violating observables can be constructed by combining the differential decay rates of $d \Gamma(\bar B \rightarrow K^- \pi^+\ell^+\ell^-)$ 
and   $d \bar{\Gamma}(B \rightarrow K^+  \pi^-\ell^+\ell^-)$.
Besides  the CP asymmetry in the dilepton mass distribution,  there are several CP violating observables
in the angular distribution.  The latter are sensitive to CP violating effects as differences between 
the angular coefficient functions, $J_i - \bar{J}_i$.  As was discussed   in Refs.~\cite{Krueger1,Krueger2}, and more recently  in Ref.~\cite{Bobeth:2008ij}, those CP asymmetries are all very small in the SM;
they  originate from the  small CP violating imaginary part of $\lambda_u = (V_{ub}V^*_{us})/(V_{tb} V^*_{ts})$. This weak phase present in the Wilson coefficient $C_9^{eff}$ is doubly-Cabbibo suppressed  and  further suppressed by the ratio of the Wilson coeficients $(3 C_1 +C_2) / C_9 \approx 0.085$.


Another remark is that the CP assymmetries related to $J_{5,6,8,9}$ can be extracted 
from $( d  \Gamma   +  d \bar{\Gamma} )$ due to the property (\ref{Jtrafos})., and thus can be determined for an untagged equal mixture of $B$ and  $\bar B$ mesons. 
This is important for the decay modes $B^0_d \rightarrow K^{*0}  ( \rightarrow K^0 \pi^0 ) \ell^+\ell^-$
and $B_s \rightarrow \phi (\rightarrow K^+K^-) \ell^+\ell^-$
but it is  less relevant for  the self-tagging mode 
$B_d \rightarrow K^{*0} (\rightarrow K^+\pi^-)\ell^+\ell^-$.

\section{QCDf/SCET framework}

The up-to-date predictions of exclusive modes are based on QCD factorization (QCDf) and its quantum field theoretical formulation, soft-collinear effective theory (SCET) 
\cite{Beneke:2001at,Beneke:2004dp}.  
The crucial theoretical observation is that  in the limit
where the initial hadron is heavy and the final meson has a large
energy \cite{large:energy:limit} the hadronic form factors can be 
expanded in the small ratios $\Lambda_{\mathrm{QCD}}/m_b$ and
$\Lambda_{\mathrm{QCD}}/E$, where $E$ is the energy of
the light meson.  Neglecting  corrections of order $1/m_b$ and $\alpha_s$, 
the seven a priori independent $B\to K^*$ form factors  
 reduce to two universal form factors $\xi_{\bot}$ and  $\xi_{\|}$
\cite{large:energy:limit,Beneke:2000wa}. 
These relations  can be strictly derived within the QCDf and SCET   approach
and  lead  to simple factorization formulae for the $B \rightarrow K^*$ form factors 
\begin{equation}
F_i = H_i \xi^P  + \phi_B \otimes T_i  \otimes \phi^P_{K^*} + O(\Lambda/m_b)\,.
\end{equation}
There is also  a similar factorization formula for the decay amplitudes.
The rationale of such  formulae is that perturbative hard kernels like $H_i$ or $T_i$ can be 
separated from process-independent nonperturbative functions like form factors $\xi^P$ or 
light-cone wave functions $\phi_i$.

However, in general we have no means to calculate $\Lambda/m_b$ corrections to
the QCDf amplitudes so they are treated as unknown corrections. This leads to a large uncertainty of theoretical predictions based on the QCDf/SCET which we will 
try to make manifest in our phenomenological analysis.
  
  The theoretical simplifications are restricted to the kinematic region
  in which the energy of the $K^*$ is of the order of the heavy quark mass,
  i.e.~$\qsq \ll m_B^2$. Moreover, the influences of very light resonances
  below 1\gevgev question the QCD factorization results in that region. In addition, the longitudinal 
  amplitude in the QCDF/SCET approach generates a logarithmic  divergence in the limit $q^2 \rightarrow 0$ indicating problems in the theoretical description below 1\gevgev \cite{Beneke:2001at}.
    Thus, we will confine our analysis of all observables to the dilepton mass
  in the range, $1\gevgev \leqslant q^2 \leqslant 6\gevgev$.

Using the discussed  simplifications  the $K^*$ spin  amplitudes 
at leading order in $1/m_b$ and $\alpha_s$ have a very
simple form:
\begin{eqnarray}
A_{\bot L,R}&=&  \sqrt{2} N m_B(1- \sh)\bigg[  
\{ (C_9^{eff} + C_9^{eff \prime}) \mp (C_{10} + C_{10}^{\prime})\}
+\frac{2\hat{m}_b}{\sh} ( C^{eff}_{7} + C^{\prime eff}_{7} ) 
\bigg]\xi_{\bot}(E_\kstar), \nonumber  \\
A_{\| L,R}&=& -\sqrt{2} N m_B (1-\sh)\bigg[
\{ (C_9^{eff} - C_9^{eff \prime}) \mp (C_{10} - C_{10}^{\prime}) \}
+\frac{2\hat{m}_b}{\sh}( C^{eff}_{7} -  C^{\prime eff}_{7} ) \bigg] \xi_{\bot}(E_\kstar)\, , \nonumber \\
A_{0L,R}&=& -\frac{Nm_B }{2 \hat{m}_\kstar \sqrt{\sh}} (1-\sh)^2\bigg[ \{ (C_9^{eff} - C_9^{eff\prime})  \mp (C_{10} - C_{10}^{\prime}) \}
+ 2
\hat{m}_b (C^{eff}_{7} -  C^{'eff}_{7}) \bigg]\xi_{\|}(E_\kstar)\, , \nonumber \\
A_t  &=& \frac{Nm_B }{ \hat{m}_\kstar \sqrt{\sh}} (1-\sh)^2\bigg[  C_{10} - C_{10}^{\prime} \bigg] \xi_{\|}(E_\kstar)
\end{eqnarray}
with $\sh =  \qsq/m_B^2$, $\hat{m}_i =  m_i/m_B$. Here we neglected  
terms of $O(\hat{m}_{K^*}^2)$.

Most  recently, a QCDf/SCET   analysis of  the angular  CP violating observables,
based on the NLO results  in Ref.  \cite{Beneke:2001at}, 
was presented for the first time~\cite{Bobeth:2008ij}.
The NLO corrections are shown to be 
sizable. The crucial impact of the NLO analysis is that the scale dependence gets reduced  to the $10\%$ level for most of the CP asymmetries. However, for some of them, which essentially start
with a nontrivial  NLO  contribution,  there is a significantly larger scale dependence.
The $q^2$-integrated  SM predictions are all shown to be below  the $10^{-2}$ level due to the small
weak phase as mentioned above. The uncertainties due to the
form factors, the scale dependence, and the uncertainty due to CKM parameters are identified 
as the main sources of SM errors~\cite{Bobeth:2008ij}. 

\section{New physics  reach}

The new physics sensitivity of  CP violating  observables in the mode $\bar B_d \rightarrow \bar K^* \ell^+ \ell^-$ was  discussed   in a model-independent way \cite{Bobeth:2008ij}  and also in various popular concrete NP models \cite{Altmannshofer:2008dz}. 
It was found that the NP contributions to the phases of the Wilson coefficients $C_7$, $C_9$, and 
$C_{10}$  and of their chiral counterparts  drastically enhance
such CP violating observables, while presently most of those phases are  very weakly constrained.  It  was claimed  that these observables offer clean signals  of NP contributions.

However, we show that the NP reach of such observables can only be  judged with a {\it complete}
analysis of the theoretical and experimental uncertainties. All details 
of our analysis with further results will be published in a forthcoming 
paper~\cite{Egedeetal2}. Here we  will restrict ourselves to the most important  issues. 
To the very detailed analyses in Refs.~\cite{Bobeth:2008ij,Altmannshofer:2008dz} 
 we add the following crucial points:
\begin{itemize} 
\item  We redefine the various CP  asymmetries following the general method presented in our
 previous paper \cite{Egede:2008uy}: An appropriate normalisation of the CP asymmetries 
 almost eliminates any uncertainties due to the soft form factors which is one of the major sources 
 of errors in the SM prediction. 
 \item We explore the effect of the possible $\Lambda/m_b$ corrections and make the uncertainty due  to those  unknown $\Lambda/m_b$ corrections manifest 
 in our analysis within  the SM and NP scenarios. 
  \item We investigate the experimental sensitivity of the angular CP assymmetries using a toy Monte Carlo model and estimate the statistical uncertainty of the observables with statistics correpsonding to
 five years of nominal running at LHCb ($10 fb^{-1}$) using a full angular fit method.
  \end{itemize}
We discuss these issues by example  the two  angular asymmetries 
corresponding to the angular coefficient functions $J_{6 s}$ and $J_8$;
\begin{equation}
 A_{6s}= \left( J_{6s} - \bar J_{6s} \right) / d(\Gamma+\bar\Gamma)/dq^2, \quad
  A_{8} = \left( J_{8} - \bar J_{8} \right) / d(\Gamma+\bar\Gamma)/dq^2
\label{definitions}
\end{equation}
`Within the SM  the first CP asymmetry  related to   $J_{6s} $  turns out to be  the well-known 
forward-backward CP asymmetry which was proposed in Refs.~\cite{G*3,Krueger3}.

\begin{figure}\includegraphics[width=.5\textwidth]{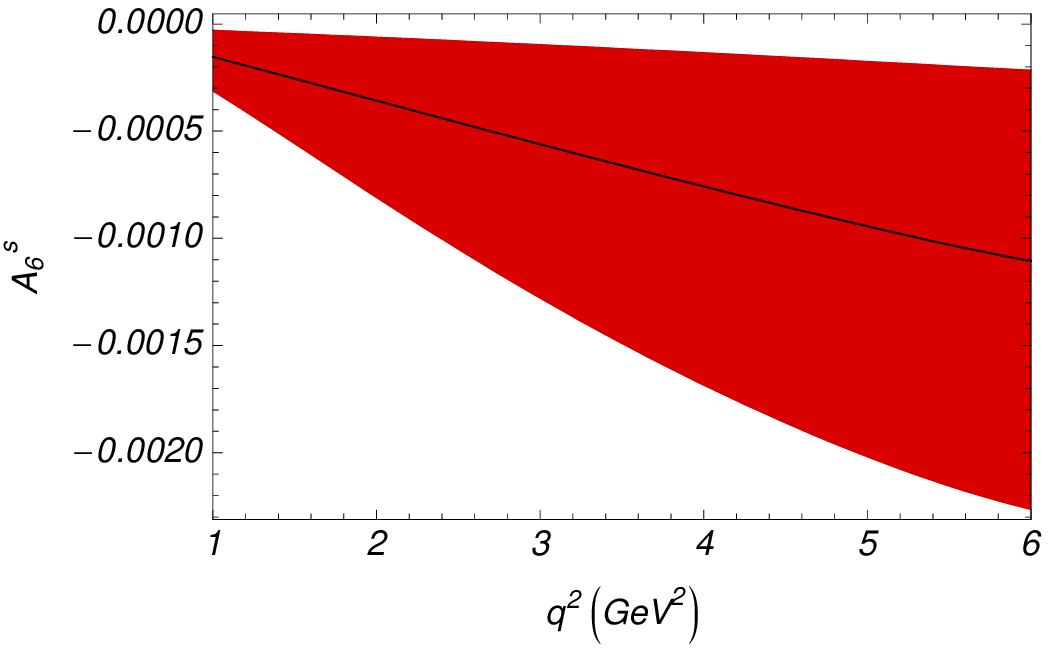}\includegraphics[width=.5\textwidth]{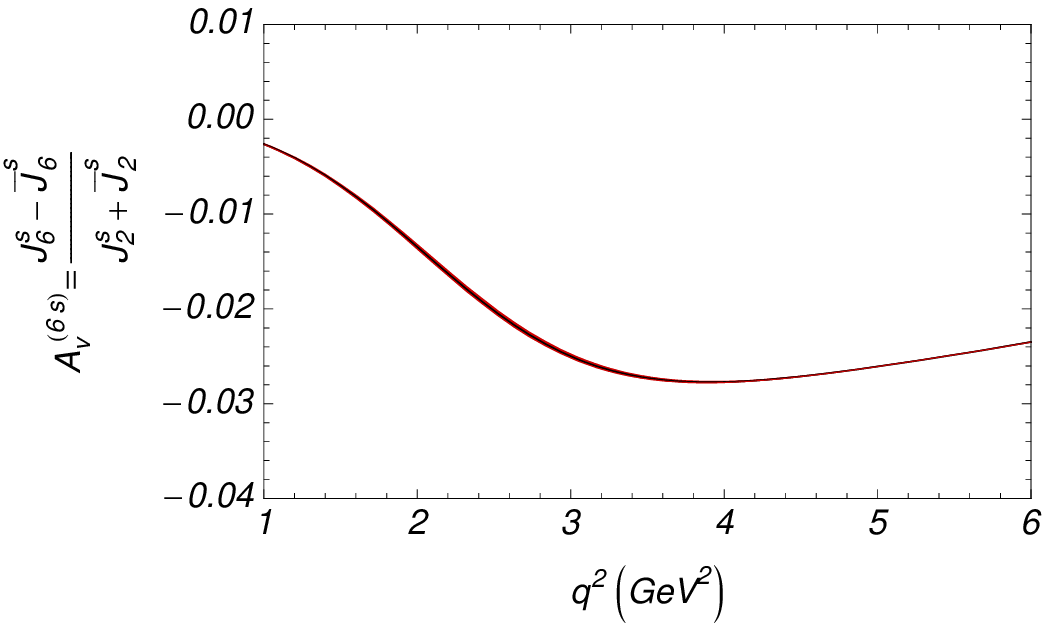}
\caption{SM prediction    of the CP violating observables $A_{6s}$ (left) and 
$A_{6s}^{V2s}$ (right)
                as function of the squared lepton mass with uncertainty due to the soft form factors only.}
\label{formfactorA62s}
\end{figure}

\begin{figure}\includegraphics[width=.5\textwidth]{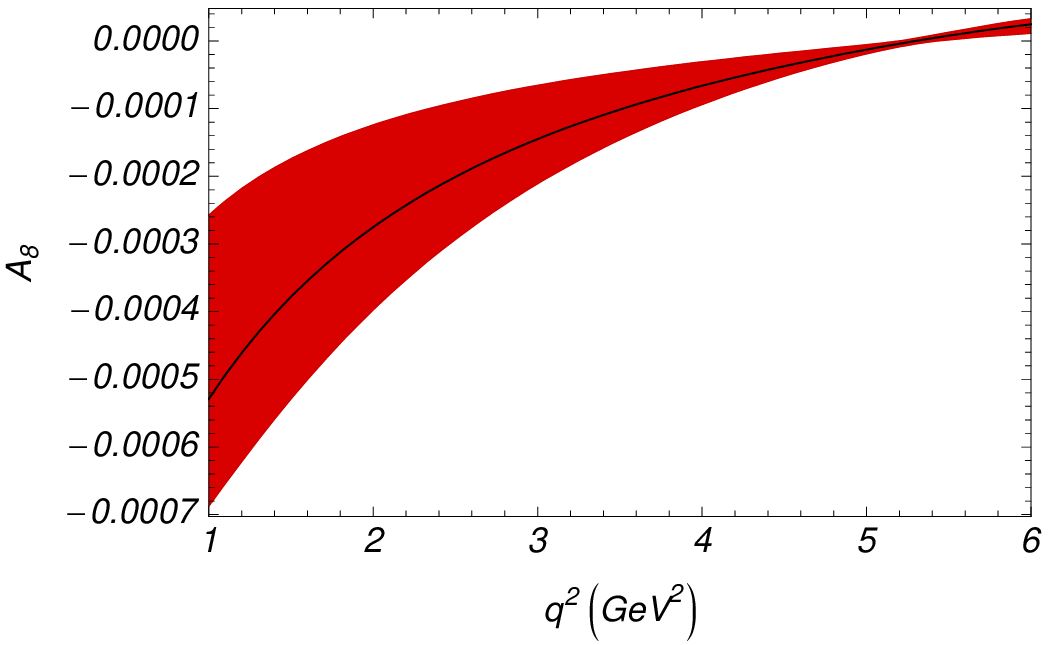}\includegraphics[width=.5\textwidth]{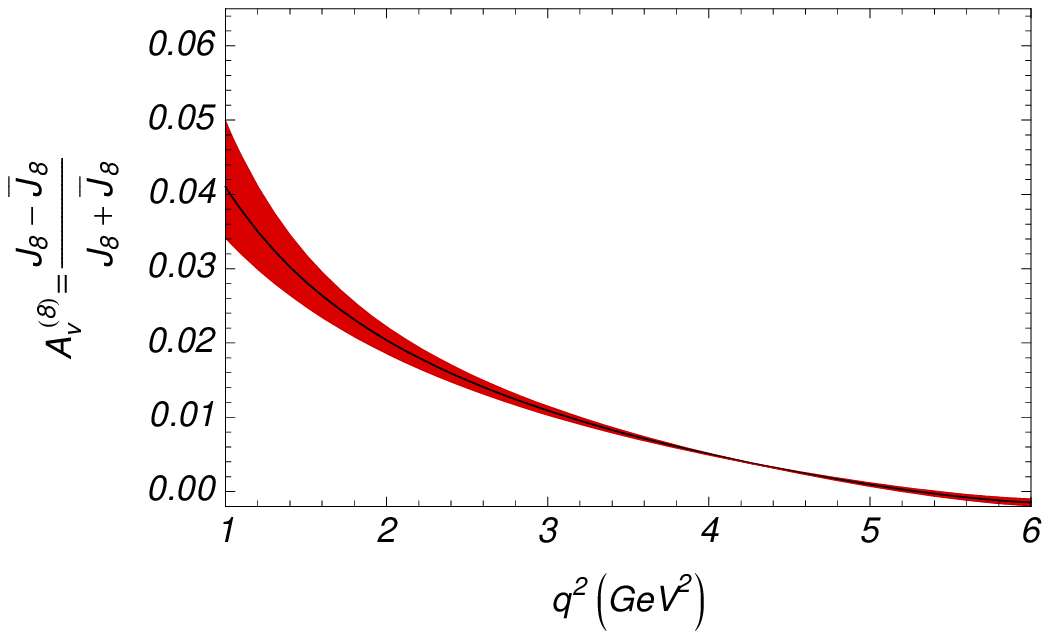}
\caption{SM prediction of  the CP violating observables $A_{8}$ (left) and 
$A_{8}^V$ (right)    with uncertainty due to the soft form factors only.}      
\label{formfactorA8}
\end{figure}

As a first step we redefine the two CP observables. We  make  sure that the form factor dependence cancels out at the LO level by using an appropriate normalisation: 
\begin{equation}
 A^{V2s}_{6s}= \left( J_{6s} - \bar J_{6s} \right) /    \left( J_{2s} +  \bar J_{2s} \right), \quad
  A^V_{8} = \left( J_{8} - \bar J_{8} \right) /  \left( J_{8} + \bar J_{8} \right)
\label{newdefinitions}
\end{equation}
The $J_i$ are bilinear in the $K^*$ spin amplitudes, so it is clear from the LO formulae  (\ref{SCETKspin})
 that -following the strategy of Ref.\cite{Egede:2008uy} - any form factor dependence at this order cancels out in both observables. We note that $J_{2s}$ has 
the same form factor dependence as $J_{6s}$ but has larger absolute values over the dilepton mass spectrum  that  stabilizes the quantity.  
In Fig.~\ref{formfactorA62s} the uncertainty due to the form factor dependence is estimated in a
conservative way (for more details see Ref.~\cite{Egedeetal2}) for $A_{6s}$ defined in 
(\ref{definitions}) and for $A_{6s}^V$ defined in (\ref{newdefinitions}). Comparing the plots, one sees that 
with the appropriate normalization this main source of hadronic uncertainties gets almost eliminated.  
The left-over uncertainty enters through the form factor dependence of the NLO contribution. 
Fig.~\ref{formfactorA8}  shows the analogous results for the observable $A_8^V$.

In the second step we try to make the possible $\Lambda/m_b$ corrections manifest in our final results.
To explore the corresponding uncertainties  we introduce a set of extra parameters for each spin 
amplitude:
\begin{equation}
\label{power1}
A = A_{1} \times (1 + C_1 e^{i\phi_1}) + e^{i\theta} A_{2} \times (1 + C_2 e^{i\phi_2}),
\end{equation}
where $A_{1,2}$ are the relevant sub--amplitudes and $\theta$ is the weak phase. We assume that the subamplitudes
can each receive a $\Lambda/m_b$ correction as well as an additional, currently unconstrained strong phase, $\phi_{1,2}$. For the absolute size of the corrections we use a dimensional estimate fixing
$C_{1,2}$ to be of order $5-10\%$. 
To access the effects of these
uncertainties on the individual observables, we form an ensemble of theory
 predictions, where each amplitude is randomly
assigned values of $\phi_{1,2}$ and $C_{1,2}$ from a Uniform distribution over the specified ranges. It is assumed that
the values of these 
parameters are not functions of $q^2$. This ensemble is used to calculate a 66\% confidence
band for each observable by looking at the spread of predictions for each 
observable at each point in $q^2$. The bands
produced show the expected uncertainty on each observable given the 
estimated ranges for the unknown parameters.

\begin{figure}\includegraphics[width=.5\textwidth]{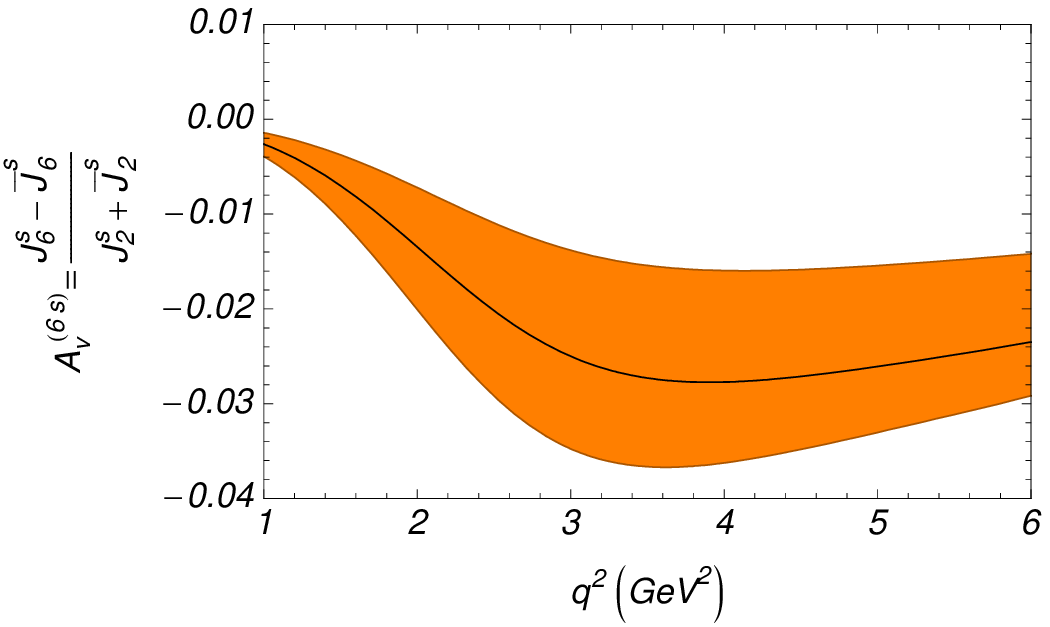}\includegraphics[width=.5\textwidth]{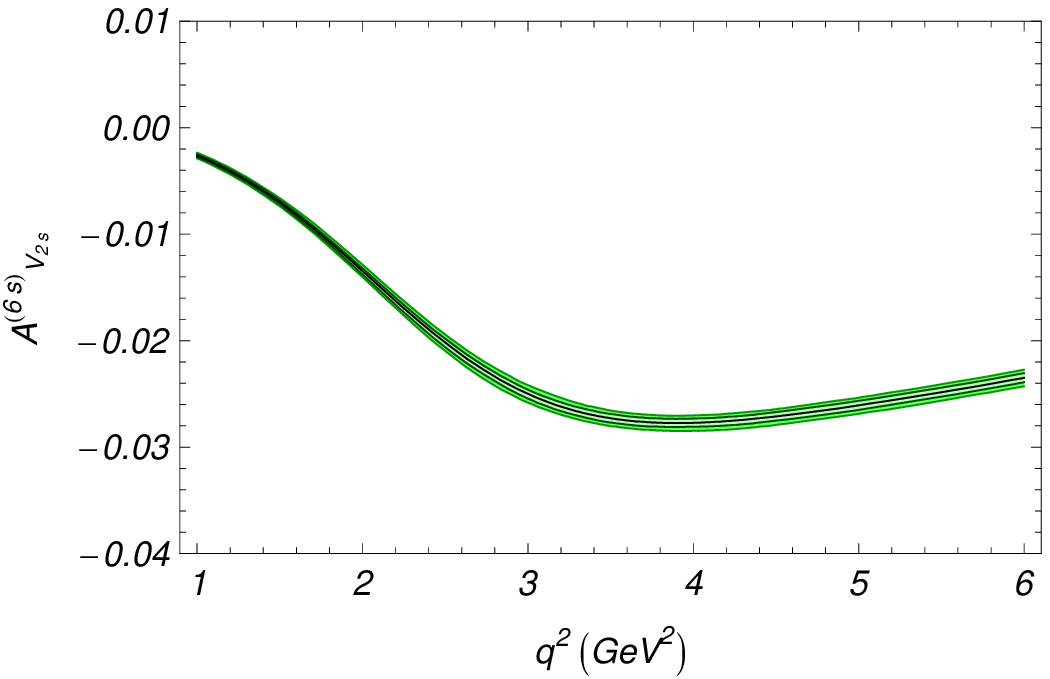}
\caption{SM uncertainty in $A_{6s}^{V2s}$ (left)  and estimate of uncertainty due to $\Lambda/m_b$ corrections with $C_{1,2} = 10\%$ (right).}
\label{SMerrorA6s}
\end{figure}

\begin{figure}\includegraphics[width=.5\textwidth]{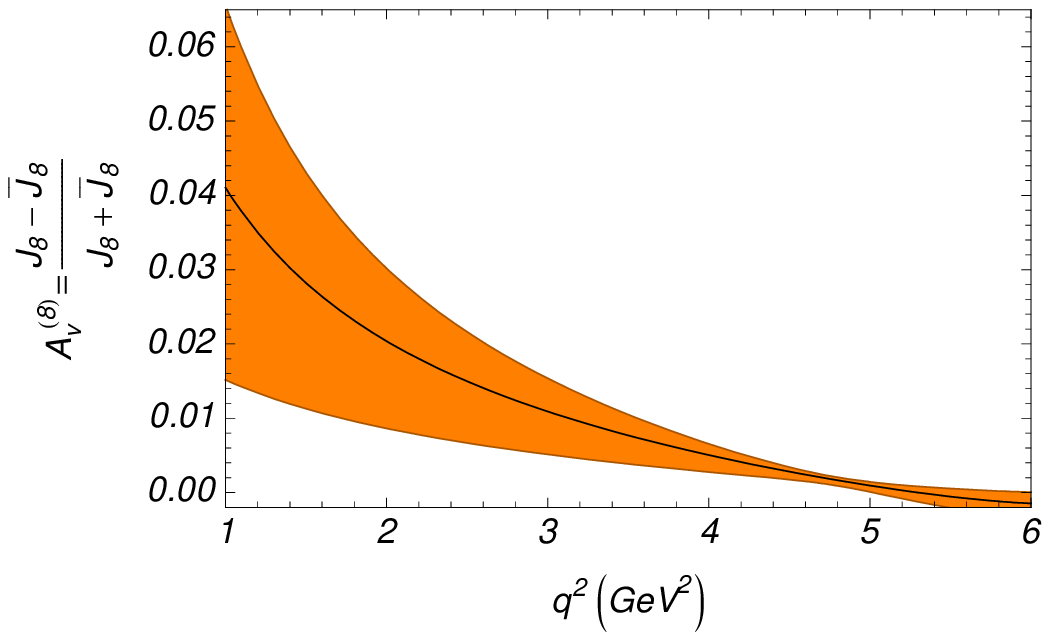}\includegraphics[width=.5\textwidth]{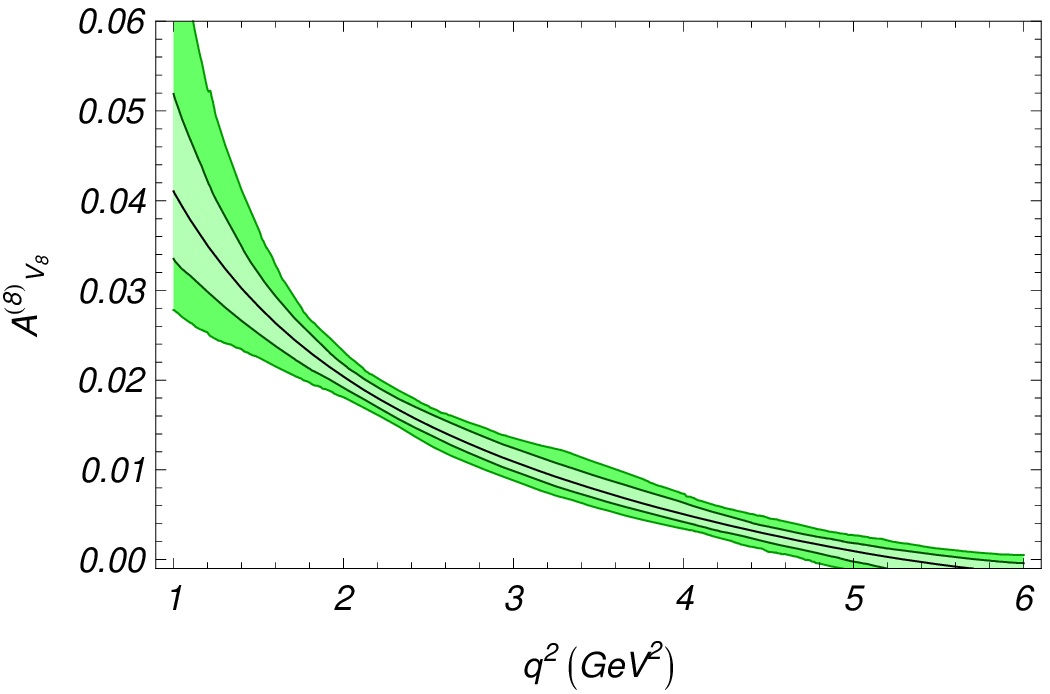}
\caption{SM uncertainty in $A_{8}^{V}$  (left) and estimate of uncertainty due to $\Lambda/m_b$ corrections (right, light grey  (green) corresponds to $C_{1,2} = 5\%$, dark grey (green) to $C_{1,2} = 10\%$).}
\label{SMerrorA8}
\end{figure}

Within the SM,  we have only one weak phase and the two subamplitudes
are contructed in the following way;
\begin{equation}
A   = A_{\rm SM} (\lambda_u=0) \times  (1+C_1 e^{i\phi_1}) + (A_{\rm SM} (\lambda_u \neq 0) - A_{\rm SM} (\lambda_u = 0)) \times (1+ C_2  e^{i \phi_2})
\label{power2}
\end{equation}
It turns out that in spite of this very conservative ansatz for the possible power corrections - we neglect
for example any kind of correlations between such corrections in the various spin amplitudes - 
the impact of those corrections is smaller than the SM uncertainty in case of the two observables 
$A_{6s}^V$ and $A_8^V$. In the left plot of Fig.~\ref{SMerrorA6s} the SM error is given including
uncertainties due to the scale dependence and  input parameters and the spurious error due to the 
form factors. 
In the right plot the estimated power corrections are given, which in case of the CP violating
observable $A_{6s}^V$ are significantly smaller than the combined uncertainty due to scale 
and input parameters.  Fig.~\ref{SMerrorA8} shows the same feature for the CP violating 
observable $A_8^V$. 
This result is in contrast to the one for CP-averaged angular observables discussed in Ref.~\cite{Egede:2008uy}, where the estimated power corrections always represent the dominant error. 
As the reason for this specific feature one identifies the smallness of the weak phase in the 
SM. Thus, one expects that the impact of power corrections  will be significantly larger 
when NP scenarios with new CP phases are considered (see below).

\begin{figure}\includegraphics[width=.5\textwidth]{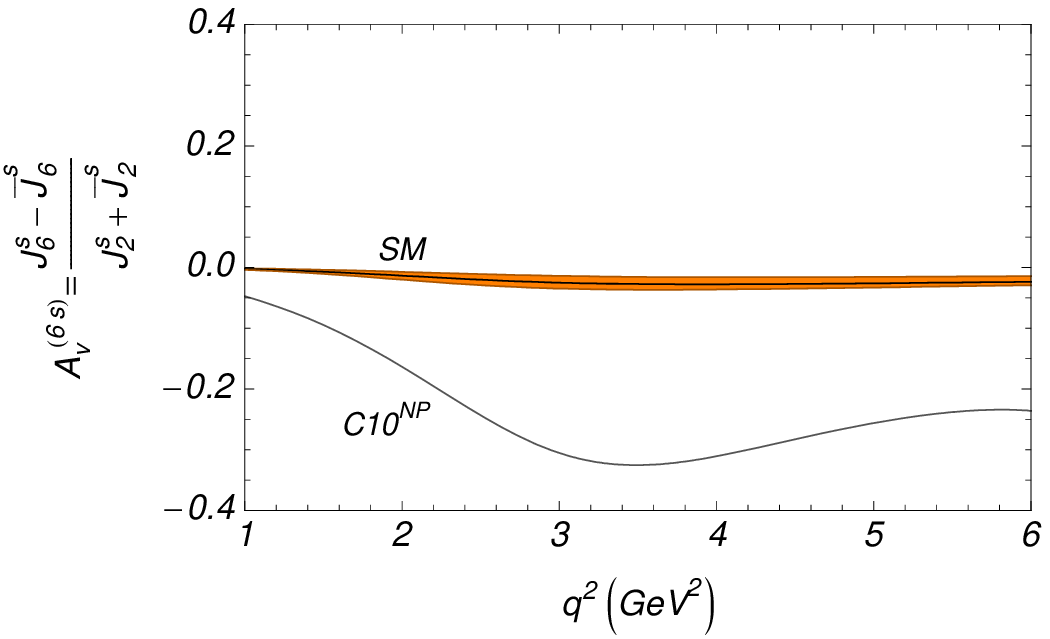}\includegraphics[width=.5\textwidth]{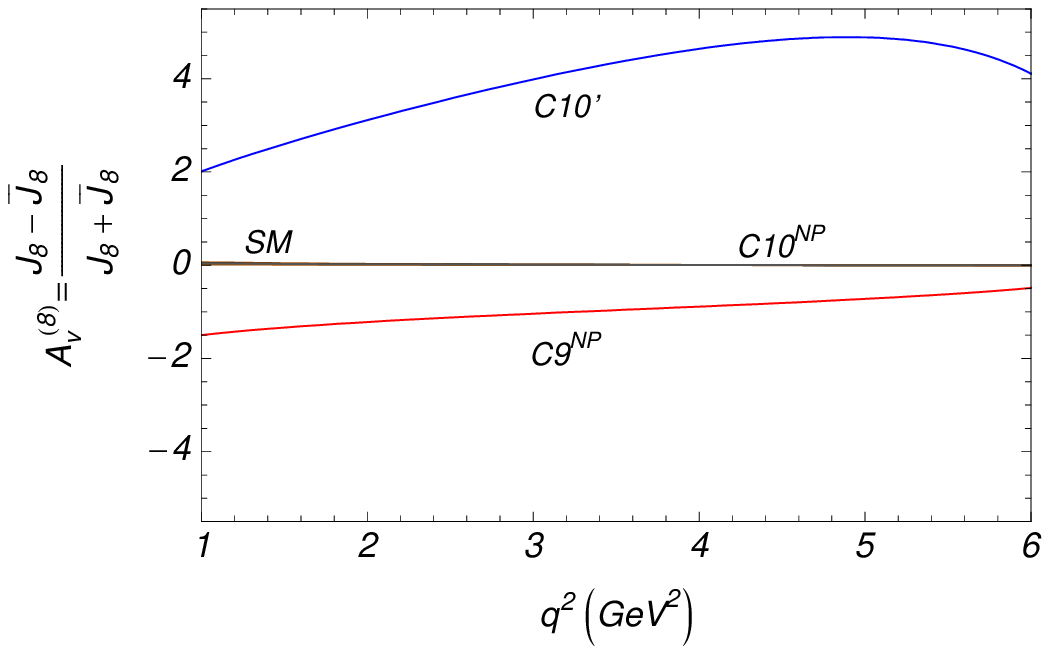}
\caption{New physics scenarios, assuming one nontrivial NP Wilson coefficient at a time, next to SM prediction for $A_{6s}^{V2s}$ (left) and $A_8^V$ (right),
for concrete values see text.}
\label{newphysics}
\end{figure}

In the third  step we consider various NP scenarios. Here we follow the model-independent constraints derived in Ref.~\cite{Bobeth:2008ij} assuming only one NP Wilson coefficient being nonzero. 
We consider three  different NP benchmarks  of this kind:
\begin{enumerate}
\item  $| C_{9}^{\rm NP} | = 2$. and $\Theta_9^{\rm  NP} = \pi/8$  
\item  $| C_{10}^{\rm NP} | = 1.5$. and $\Theta_{10}^{\rm  NP} = \pi/8$ 
\item   $| C^{'}_{10} | = 3$. and $\Theta_{10}^{'} = \pi/8$ 
\end{enumerate} 
The absolute values are chosen in such a way that the model-independent analysis,  assuming {\it one}
nontrivial NP Wilson coefficient at a time, does not give any bound on the 
corresponding NP phase. Scenarios with larger phase values will be discussed in 
Ref.~\cite{Egedeetal2}. 
Fig.~\ref{newphysics}  shows that the CP violating observable $A_{6s}^V$ might separate 
a NP scenario (2), while the central values of scenarios  (1) and (3) are very close to the SM. Moreover  observable $A_*^V$ seems to  be suited to separate scenarios (1) and (3) from the SM.

However, to judge the NP reach we need a complete error analysis within the three NP scenarios.
Thus, let us consider the possible impact of unknown power corrections in these cases: With one new 
 CP phase involved we work now with three weak subamplitudes in which possible power
 corrections are varied independently. 

\begin{figure}\includegraphics[width=.5\textwidth]{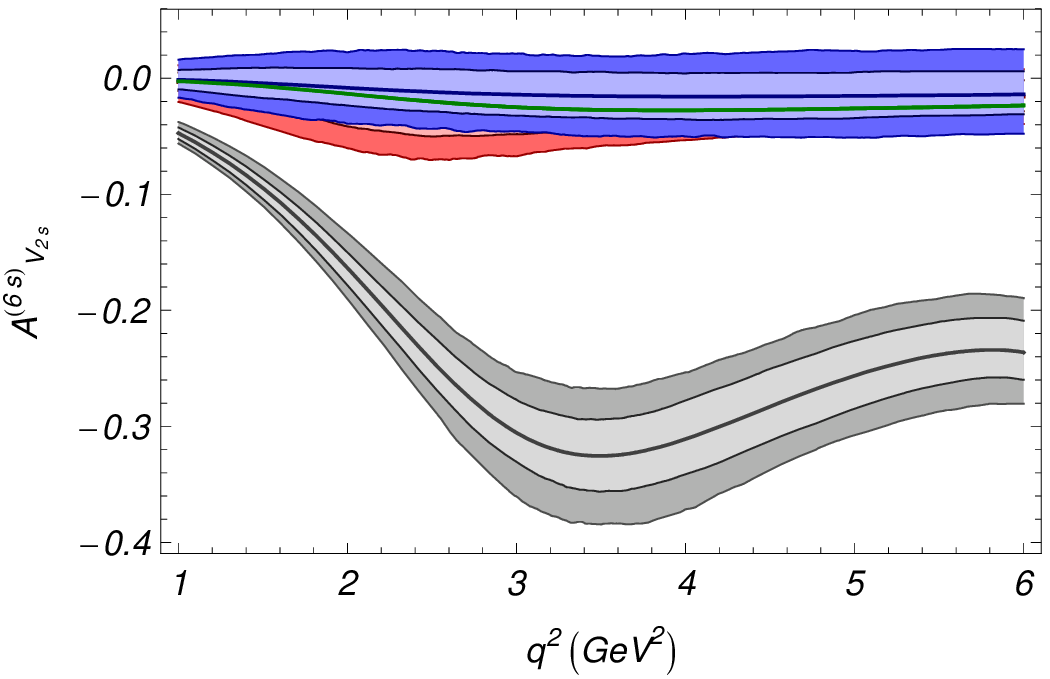}\includegraphics[width=.5\textwidth]{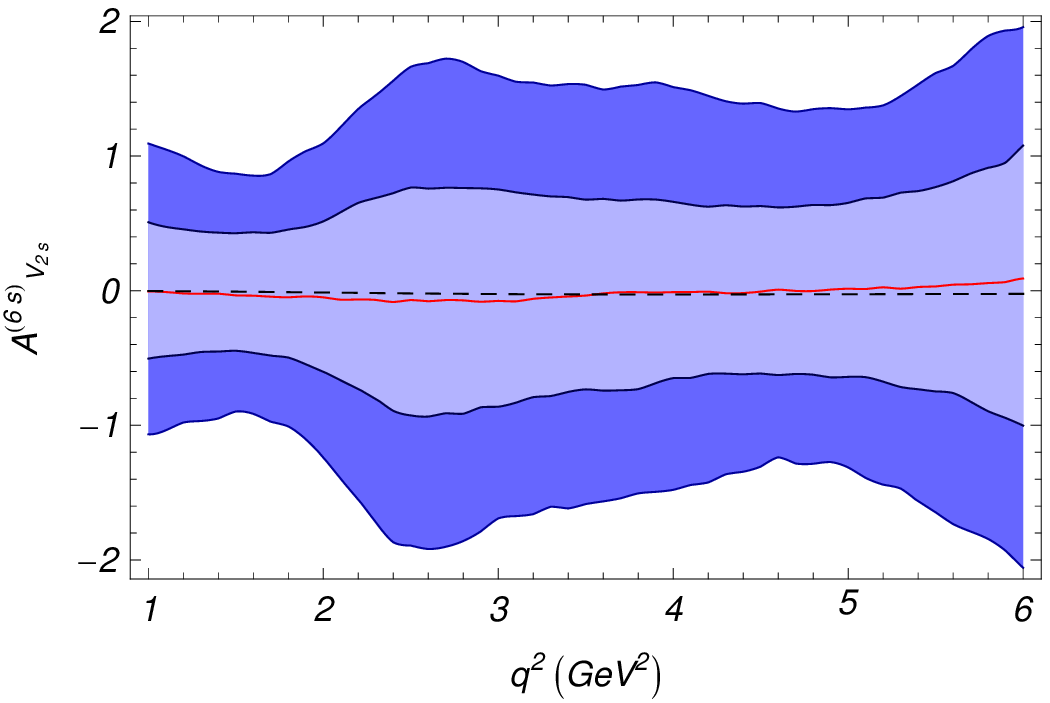}
\caption{$A_{6s}^{V2s}$: Estimate of uncertainty due to $\Lambda/m_b$ corrections (left) and experimental uncertainty (right).}
\label{lambda.exA6s}
\end{figure}

\begin{figure}
\includegraphics[width=.5\textwidth]{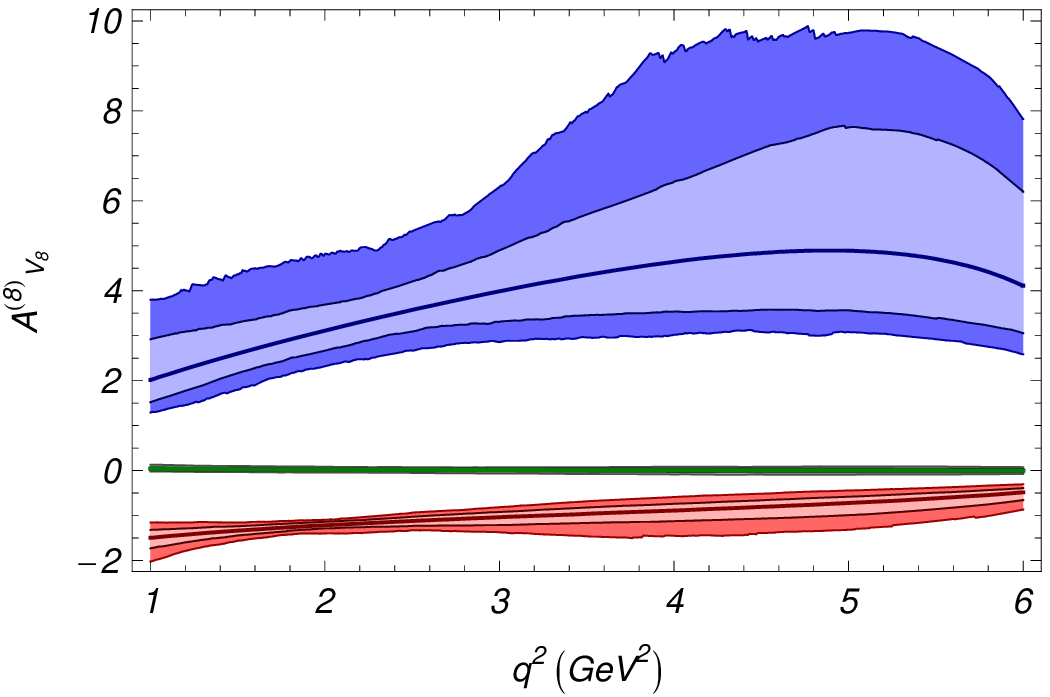}\includegraphics[width=.5\textwidth]{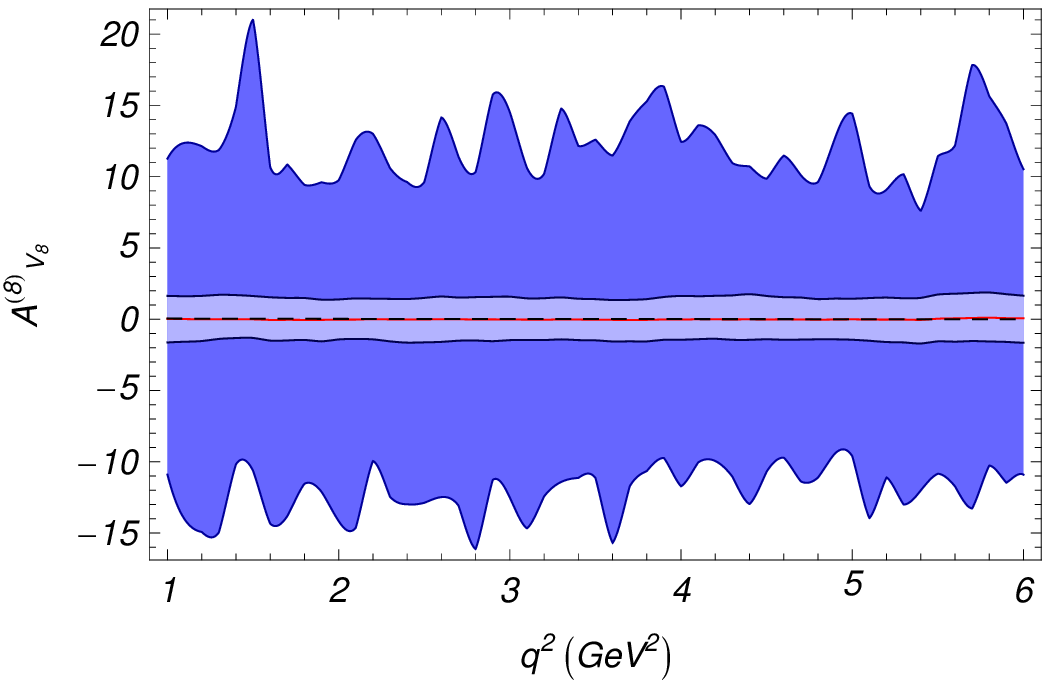}
\caption{$A_8^V$:   Estimate of uncertainty due to $\Lambda/m_b$ corrections (left) and experimental uncertainty (right).}
\label{lambda.exA8}
\end{figure}

The left plots in Figs.~\ref{lambda.exA6s} and \ref{lambda.exA8} show that the possible  $\Lambda/m_b$ corrections have a much larger impact  than in the SM and become the dominating 
theoretical uncertainty. However, the two CP violating 
observables could  discriminate  the specific NP scenarios with new CP phase of order $\pi/8$
from the SM  in view of the theoretical errors only.

In the last step, we analyse  the experimental sensitivity of the angular CP asymmetries using a toy Monte Carlo model.   The right plots in Figs.~\ref{lambda.exA6s} and \ref{lambda.exA8}  show    the estimates of the statistical uncertainty of $A_{6s}^V$ and $A_8^V$ with statistics corresponding to
 five years of nominal running at LHCb ($10 fb^{-1}$).  The inner 
and outer bands correspond to $1\sigma$  and $2\sigma$ statistical errors. 
The plots show that all the NP benchmarks are within the $1\sigma$ range of the expected
experimental error in case of the observable $A_{6s}^V$, and within the $2\sigma$ range of the
experimental error in case  of the observable $A_8^V$.

Thus, our final conclusion is that while the prospects of NP discovery of the 
CP conserving observable presented in ref.~\cite{Egede:2008uy}  both from the 
theoretical  and 
experimental point 
of view are excellent, the possibility to disentangle different NP 
scenarios for the CP violating 
observables  
remains rather 
difficult.  For the studied observables, LHCb has no real  sensitivity for NP phases up to
values of $\pi/8$  in the Wilson coefficients $C_9$, $C_{10}$, and their chiral counterparts via the rare decays $\bar B \rightarrow \bar K^* \ell^+\ell^-$. 
Even Super-LHCb with $100 fb^{-1}$ integrated luminosity does not improve the situation significantly.


\newpage

\footnotesize
\section*{Acknowledgement}
This work is supported by the European Network MRTN-CT-2006-035505 'HEPTOOLS'. TH
acknowledges \mbox{financial} support of the ITP at the University Zurich.



\begin{thebibliography}{99}

{\footnotesize

\newcommand{\np}[3]{Nucl. Phys. {\bf B#1} (#2) #3}
\newcommand{\pl}[3]{Phys. Lett. {\bf B#1} (#2) #3}
\newcommand{\pr}[3]{Phys. Rev.  {\bf D#1} (#2) #3}
\newcommand{\prl}[3]{Phys. Rev. Lett. {\bf #1} (#2) #3}
\newcommand{\prp}[3]{Phys. Rept. {\bf #1} (#2) #3}
\newcommand{\ptp}[3]{Prog. Theor. Phys. {\bf #1} (#2) #3}
\newcommand{\zpc}[3]{Z. Phys. {\bf C#1} (#2) #3}
\newcommand{\ibid}[3]{{\it ibid.} {\bf #1} (#2) #3}

\bibitem{BABAR}
http://www.slac.stanford.edu/BFROOT/

\bibitem{BELLE}
http://belle.kek.jp/

\bibitem{TevatronB1}
http://www-cdf.fnal.gov/physics/new/bottom/bottom.html

\bibitem{TevatronB2}
http://www-d0.fnal.gov/Run2Physics/ckm/


\bibitem{CERNFLAVOUR2}
  M.~Artuso {\it et al.},
  Eur.\ Phys.\ J.\  C {\bf 57} (2008) 309
  [arXiv:0801.1833 [hep-ph]].


\bibitem{CKM}
  M.~Kobayashi and T.~Maskawa,
  Prog.\ Theor.\ Phys.\  {\bf 49} (1973) 652.


\bibitem{NOBEL}
http://nobelprize.org


\bibitem{Egede:2008uy}
  U.~Egede, T.~Hurth, J.~Matias, M.~Ramon and W.~Reece,
  arXiv:0807.2589 [hep-ph].
  
  
  \bibitem{Egedeetal1b}
  U.~Egede, T.~Hurth, J.~Matias, M.~Ramon and W.~Reece,
 talk at Flavianet Meeting  \\ ``The exclusive New $B \rightarrow K^* (\rightarrow K \pi)  \ell^+ \ell^-$,'' 
Kazimierz, July 2009,  [arXiv:0912.1339 [hep-ph]].
 

  
\bibitem{Egedeetal2}
  U.~Egede, T.~Hurth, J.~Matias, M.~Ramon and W.~Reece, in preparation.


\bibitem{Krueger1}
F.~Kr\"uger, L.~M.~Sehgal, N.~Sinha, and R.~Sinha, \pr{61}{2000}{114028};
\pr{63}{2001}{019901(E)}.


\bibitem{Krueger2}
  F.~Kr\"uger, Chapter 2.17 in 
 J.~L.~.~Hewett {\it et al.},
  ``The discovery potential of a Super B Factory. Proceedings, SLAC  Workshops,
  Stanford, USA, 2003,''  arXiv:hep-ph/0503261.
 


  
\bibitem{Bobeth:2008ij}
  C.~Bobeth, G.~Hiller and G.~Piranishvili,
  JHEP {\bf 0807} (2008) 106
  [arXiv:0805.2525 [hep-ph]].


\bibitem{Beneke:2001at}
  M.~Beneke, T.~Feldmann and D.~Seidel,
  Nucl.\ Phys.\  B {\bf 612} (2001) 25
  [arXiv:hep-ph/0106067].



\bibitem{Beneke:2004dp}
  M.~Beneke, T.~Feldmann and D.~Seidel,
  Eur.\ Phys.\ J.\  C {\bf 41} (2005) 173
  [arXiv:hep-ph/0412400].

\bibitem{large:energy:limit}
  J.~Charles, A.~Le Yaouanc, L.~Oliver, O.~Pene and J.~C.~Raynal,
  Phys.\ Rev.\  D {\bf 60} (1999) 014001
  [arXiv:hep-ph/9812358].



\bibitem{Beneke:2000wa}
  M.~Beneke and T.~Feldmann,
  Nucl.\ Phys.\  B {\bf 592} (2001) 3
  [arXiv:hep-ph/0008255].



\bibitem{G*3}
  G.~Buchalla, G.~Hiller and G.~Isidori,
  Phys.\ Rev.\  D {\bf 63}, 014015 (2000)
  [arXiv:hep-ph/0006136].

\bibitem{Krueger3}
  F.~Kruger and E.~Lunghi,
  Phys.\ Rev.\  D {\bf 63}, 014013 (2001)
  [arXiv:hep-ph/0008210].


\bibitem{Altmannshofer:2008dz}
  W.~Altmannshofer, P.~Ball, A.~Bharucha, A.~J.~Buras, D.~M.~Straub and M.~Wick,
  JHEP {\bf 0901} (2009) 019
  [arXiv:0811.1214 [hep-ph]].






}


 
\end{thebibliography}
\end{document}